\documentclass[final,5p,times,twocolumn]{elsarticle}

\usepackage{amssymb}
\usepackage{amsmath}
\usepackage{lipsum}
\usepackage{lineno}
\usepackage{booktabs}  
\usepackage{makecell}   
\usepackage{graphicx} 
\graphicspath{{./Images/}} 
\usepackage{float}
\usepackage{dcolumn}
\usepackage{bm}
\usepackage{nameref} 

\newcommand{\beq}{\begin{eqnarray}} 
\newcommand{\eeq}{\end{eqnarray}}

\def\nuc#1#2{\relax\ifmmode{}^{#1}{\protect\text{#2}}\else${}^{#1}$#2\fi}

\newcommand{\vecq}{{\bm{q}}}
\newcommand{\veck}{{\bm{k}}}

\journal{Physics Letters B}
\usepackage{hyperref}
\hypersetup{
    colorlinks=true,  
    linkcolor=blue,   
    citecolor=blue,   
    urlcolor=blue,     
    pdfencoding=auto, 
    psdextra,         
}
\begin{document}

\begin{frontmatter}



\title{Implications of relativistic corrections on high-momentum nucleon-transfer reactions}




\author[label1]{W.~L.~Hai} 

\author[label1]{D.~Y.~Pang\corref{cor1}}
\cortext[cor1]{Corresponding authors}
\ead{dypang@buaa.edu.cn}  

\author[label2,label1]{I.~Tanihata}

\author[label3,label4,label5,label2,label6]{H.~J.~Ong\corref{cor1}}
\ead{onghjin@impcas.ac.cn}  

\author[label3,label2,label6]{S.~Terashima}

\author[label1,label2]{X.~Wang}

\author[label7]{Y.P.~Xu}

\author[label8]{W.D.~Chen}

\author[label1]{R.Y.~Chen}

\author[label1]{J.J.~Yan}

\affiliation[label1]{
  organization={School of Physics, Beihang University},
  city={Beijing},
  postcode={100191},
  country={China}
}

\affiliation[label2]{
  organization={Research Center for Nuclear Physics, Osaka University},
  city={Osaka},
  postcode={567-0047},
  country={Japan}
}

\affiliation[label3]{
  organization={State Key Laboratory of Heavy Ion Science and Technology, Institute of Modern Physics, Chinese Academy of Sciences},
  city={Lanzhou},
  postcode={730000},
  country={China}
}

\affiliation[label4]{
  organization={School of Nuclear Science and Technology, University of Chinese Academy of Sciences},
  city={Beijing},
  postcode={100049},
  country={China}
}

\affiliation[label5]{
  organization={Joint Department for Nuclear Physics, Lanzhou University and Institute of Modern Physics, Chinese Academy of Sciences},
  city={Lanzhou},
  postcode={730000},
  country={China}
}

\affiliation[label6]{
  organization={Nishina Center for Accelerator-Based Science, RIKEN},
  city={Saitama},
  postcode={351-0198},
  country={Japan}
}

\affiliation[label7]{
  organization={School of Nuclear Science and Engineering, North China Electric Power University},
  city={Beijing},
  postcode={102206},
  country={China}
}

\affiliation[label8]{
  organization={Institute of Applied Physics and Computational Mathematics},
  city={Beijing},
  postcode={100094},
  country={China}
}

\corref{cor1}


\begin{abstract}
High-momentum components (HMCs) of nuclear wave functions, governed by short-range nucleon-nucleon correlations, provide essential insights into nuclear structure beyond the mean-field picture. 
High-energy $(p,d)$ reactions offer access to these HMCs, but their theoretical treatment requires relativistic corrections when incident proton energies reach several hundred MeV. Although effects of relativistic kinematic corrections (RKCs) have been studied in several types of direct nuclear reactions, it has not been systematically studied in nucleon transfer reactions.
Here, RKCs are incorporated into the adiabatic distorted wave approximation (ADWA) for $(p,d)$ reactions by redefining particle masses in the zero-momentum frame. 
The approach is validated against proton elastic scattering data on \nuc{16}{O} from 135 to 800 MeV using Dirac global optical model potentials, and then applied to ($p$,$d$) reactions on \nuc{12}{C}, \nuc{16}{O}, and \nuc{40}{Ca} at incident energies from approximately 50 to 800 MeV. 
The RKCs yield neutron spectroscopic factors that are significantly more consistent across the entire energy range than those obtained from non-relativistic calculations, which systematically overestimate spectroscopic factors obtained at high incident energies. 
The present analysis demonstrates that relativistic kinematic corrections are of fundamental importance for the reliable extraction of spectroscopic factors and the accurate description of high-momentum nucleon-transfer reaction data.
\end{abstract}




\end{frontmatter}



Models of nuclear interactions and the description of atomic nuclei based on these interactions are central topics of nuclear physics. 
Unlike the low-momentum components of nuclear wave functions, which can be rather well described by sophisticated mean-field potentials, the high-momentum components (HMCs) are governed by short-range nucleon-nucleon interactions. 
These interactions manifest as short-range correlations -- such as those induced by the tensor interactions -- among the constituents of the atomic nucleus~\cite{Subedi2008}. Describing these HMCs requires nuclear theory to go beyond the mean-field framework. Consequently, experimental and theoretical studies of HMCs have remained a central theme in nuclear physics for several decades~\cite{Subedi2008, Sick-PRL-1980, Ciofi-PRC-1996, Sick-PPNP-2007,  Arrington-PPNP-2012, Hen-Science-2014, CLAS-nature-2018, Schmidt-nature-2020}.

The HMCs of nuclear wave functions can be probed using high-energy electron-induced nucleon-knockout reactions, including inclusive ($e$,$e'p$) reactions~\cite{Sick-PRL-1980, Ciofi-PRC-1996, Sick-PPNP-2007, Arrington-PPNP-2012}. While numerous such measurements have been performed on stable nuclei, electron scattering on radioactive nuclei remains extremely challenging~\cite{Suda2017}. In such cases, hadronic probes such as nucleon-knockout~\cite{Mardor1998,Tang2003,Patsyuk2021} and nucleon-transfer reactions~\cite{Ong2013, Terashima2018, Wang2026} offer viable alternatives. 
However, nucleon-knockout reactions, such as ($p$,$pn$) and ($p$,$2p$) reactions, are often limited by low statistics due to the requirement for coincidence detection of multiple outgoing particles, which hinders studies of individual nuclear states.

Nucleon transfer reactions -- such as the A$(p,d)$B reaction where a proton projectile picks up a neutron from the target nucleus A forming a deuteron and leaving the residual nucleus B -- are among the most widely used tools to study the single-particle structure of atomic nuclei~\cite{Austern-book,Schiffer-PRL-2012, Liu-NST-2020, KongWJ-NST-2023}. Within the non-relativistic center-of-mass framework, momentum conservation requires that
\begin{equation}\label{eq-mom-consv}
 \vecq=\veck_d-\frac{m_{\mathrm{A}}-m_p}{m_{\mathrm{A}}}\veck_p,
\end{equation}
where $\vecq$ is the momentum of the picked-up neutron in the nucleus A (rest mass $m_{\mathrm{A}}$), $m_p$ is the proton rest mass, and $\veck_p$ and $\veck_d$ are the momenta of the incident proton and the outgoing deuteron, respectively. In this work, we adopt natural units with the reduced Planck constant $\hbar$ and the speed of light $c$ set to unity ($\hbar$=$c$=1) unless otherwise stated. Energy conservation leads to the following relation between the absolute values of the momenta $k_p$ and $k_d$:
\begin{equation}
\frac{k_d^2}{2\mu_d} = \frac{k_p^2}{2\mu_p}+Q,
\label{eq-energy-consv}
\end{equation}
where $Q$ is the $Q$-value of the A$(p,d)$B reaction. The reduced masses in the incident- and outgoing-channels are $\mu_p=m_pm_{\mathrm{A}}/(m_p+m_{\mathrm{A}})$ and $\mu_d=m_dm_{\mathrm{B}}/(m_d+m_{\mathrm{B}})$, respectively, with $m_d$ and $m_{\mathrm{B}}$ denoting the rest masses of the deuteron and residual nucleus B, respectively. The proton momentum $k_p$ is related to the incident laboratory energy $T_{\mathrm{in}}^{(\mathrm{L})}$ by 
\begin{equation} \label{eq-kp-elab}
\frac{k_p^2}{2\mu_p}=\frac{m_{\mathrm{A}}}{m_{\mathrm{A}}+m_p}T_{\mathrm{in}}^{(\mathrm{L})}.
\end{equation}
Taking the direction of the incident proton as the $z$-axis of the coordinate system, the magnitude of the momentum of the picked-up neutron in nucleus A is:
\begin{equation}\label{eq-q}
q = \sqrt{k_d^2 + \left( \frac{m_{\mathrm{A}} - m_p}{m_{\mathrm{A}}} k_p \right)^2 - 2 \frac{m_{\mathrm{A}} - m_p}{m_{\mathrm{A}}} k_p k_d \cos\theta},
\end{equation}
where $\theta$ is the scattering angle of the outgoing deuteron in the center-of-mass system.

Equations~(\ref{eq-mom-consv}-\ref{eq-q}) 
suggest that the momentum of the picked-up neutron in the nucleus A can be obtained by measuring the momentum of the outgoing deuteron at a scattering angle $\theta$ in ($p$,$d$) reactions.
For a fixed incident energy $T_{\mathrm{in}}^{(\mathrm{L})}$, the momentum $q$ increases with $\theta$. Therefore, to probe HMCs of the nuclear wave functions using $(p,d)$ reactions, one can either measure the deuterons, thus the differential cross sections, at large angles with relatively low incident energies, or at small angles with high incident energies. However, the reaction mechanisms governing ($p$,$d$) reactions at large scattering angles are typically more complicated than those at small angles. 
Consequently, to minimize the model dependence in extracting the HMC information, it is preferable to perform ($p$,$d$) reaction measurements at small angles and at high incident energies.

High-momentum transfer ($p$,$d$) reactions have been proposed to probe HMCs in atomic nuclei, particularly those induced by the tensor interactions~\cite{Tanihata-MPLA-2010}. Theoretically, tensor-induced HMCs are expected to dominate at nucleon momenta around 2 fm$^{-1}$~\cite{Schiavilla-PRL-2007,Roth-PPNP-2010}, which necessitates zero-degree $(p,d)$ reaction measurements at incident energies up to approximately 800 MeV.
Indeed, experiments using proton beams at incident energies between 198 and 1209 MeV have reported observations of effects stemming from the tensor interactions in \nuc{16}{O}~\cite{Ong2013, Wang2026}. 
Theoretical description of the $(p,d)$ reactions at high energies, however, remains challenging. 
One of the main difficulties is that relativistic effects become significant and must be accounted for. To date, most direct reaction models are grounded in non-relativistic scattering theory, and, to our knowledge, no rigorous relativistic direct reaction theory has yet been proposed. 
Practically, one must therefore resort to relativistic corrections to existing reaction theories and models, among which, kinematic relativistic corrections are the most commonly applied~\cite{Ingemarsson-PC-1974, Satchler-NPA-1992, Kyushu, Pang-PRC-2009, Pang-CPC-2014}. 
In addition, the choice of reaction model is also critical. 
Although the distorted wave impulse approximation (DWIA) has been successfully applied to knockout reactions at intermediate energies, a recent comparative study for the \nuc{16}{O}($p$,$d$)\nuc{15}{O} reaction at 200 MeV with DWIA and the distorted wave Born approximation (DWBA) -- both including relativistic corrections -- reported that the DWBA provides a more suitable description of $(p,d)$ reactions~\cite{Shim2026}.

In this paper, we study $(p,d)$ reactions on \nuc{40}{Ca}, \nuc{16}{O} and \nuc{12}{C} at incident energies from 45.4 MeV to 800 MeV. We examine the effects of relativistic kinematic corrections on the descriptions of the experimental data and on the extracted single-neutron spectroscopic factors. To account for deuteron breakup effects, we employ the adiabatic distorted wave approximation (ADWA)~\cite{Johnson-PRC-1970}.


The transition amplitude for the A($p$,$d$)B reaction in the prior-form ADWA is given by \cite{Johnson-aip-2005,Timofeyuk-ppnp-2020}:
\begin{equation}
T_{pd} = S^{1/2}_{I_{\mathrm{A}}I_{\mathrm{B}},n\ell j} \langle \phi_d \chi_d^{\mathrm{(-)}} | V_{pn}| \phi_{n\ell j} \chi_p^{\mathrm{(+)}} \rangle,
\end{equation}
where $\phi_d$ is the deuteron ground-state wave function, which is calculated with the $p$-$n$ interaction $V_{pn}$. $\phi_{{n\ell j}}$ is the single-particle neutron wave function in the target nucleus A with $n$, $\ell$ and $j$ being its principal quantum number, orbital angular momentum and total angular momentum numbers, respectively. $\chi_p^{\mathrm{(+)}}$ and $\chi_d^{\mathrm{(-)}}$ are the distorted waves in the incident and outgoing channels, respectively, with the superscripts $\pm$ indicating their boundary conditions. Within the ADWA framework, $\chi_d^{\mathrm{(-)}}$ is calculated with an effective deuteron potential, which is the sum of proton and neutron optical model potentials on the same target nucleus evaluated at half of the deuteron incident energy~\cite{Johnson-aip-2005}. $S_{I_{\mathrm{A}}I_{\mathrm{B}},n\ell j}$ is the spectroscopic factor (SF), and $I_{\mathrm{A}}$ and $I_{\mathrm{B}}$ are the spins of the nuclei A and B, respectively.
Theoretically, SFs depend on the internal wave functions of the mother and daughter nuclei~\cite{Austern-book}, and should therefore be energy-independent. In real experiments and analyses, however, extracted SFs often show energy dependence due to factors such as reaction model approximations, relativistic effects, optical model uncertainties and breakdown of single-particle picture.

The distorted waves $\chi_p$ and $\chi_d$ satisfy the following equation:
\begin{equation}
    \left(\nabla^2 + {2\mu E} - {2\mu U(r)}\right)\chi = 0,
    \label{eq-dw}
\end{equation}
where $E$ is the kinetic energy in the center-of-mass (c.m.) system, $U(r)$ is the optical model potential (OMP), and $\mu$ is the reduced mass. Within the framework of non-relativistic kinematics, $\mu = m_{\mathrm{A}}m_{\mathrm{a}}/(m_{\mathrm{A}}+m_{\mathrm{a}})$ for the incident channel, and $\mu = m_{\mathrm{B}}m_{\mathrm{b}}/(m_{\mathrm{B}}+m_{\mathrm{b}})$ for the outgoing channel, where $m_i$ denotes the rest masses of particle i (i$=$A, a, B and b). With this definition, the magnitude of the relative momentum between two particles in their center-of-mass frame is $k=\sqrt{2\mu E}$.
Equation~(\ref{eq-dw}) can therefore be rewritten as 
\begin{equation}
     \left(\nabla^2 + {k^2} - {2\mu U(r)}\right)\chi = 0.
     \label{eq-dw2}
\end{equation}
At high incident energies, when $E$ becomes much larger than the depth of the potential $U$, the $k^2$ term dominates. In this regime, accurate determination of $k$ is crucial for solving the distorted wave equations and for calculating transfer reaction cross sections.

Equations (\ref{eq-energy-consv}--\ref{eq-dw2}) are formulated within non-relativistic kinematics, where reduced masses arise from separating the center-of-mass motion from the relative motion of the two colliding particles~\cite{Yabana2003StructureAR}. However, such a separation of motions is impossible at high energies, since the velocity of a particle can no longer be expressed as the sum of the velocity of the center-of-mass and its velocity relative to the center-of-mass. In relativistic cases, it is convenient to adopt a center-of-momentum frame of reference -- also known as the zero-momentum frame (ZMF) -- instead of the center-of-mass frame. 

In non-relativistic direct nuclear reaction codes, the relative momentum $p$ between two colliding particles -- projectile a and target A -- is evaluated as $p=\sqrt{2\mu E_\textrm{c.m.}}$, where $E_\textrm{c.m.}=(m_\textrm{A}/(m_\textrm{A}+m_\textrm{a}))T_\textrm{in}^\textrm{(L)}$ is the center-of-mass energy, with $T_\textrm{in}^\textrm{(L)}$ the incident laboratory energy of particle a, and $\mu=m_\textrm{A}m_\textrm{a}/(m_\textrm{A}+m_\textrm{a})$ the reduced mass. To extend the use of these codes for transfer reactions at intermediate and high energies, the particle masses and incident energies must be modified so that the kinetic energies in the incident and outgoing channels are calculated consistently, following the same prescription as in the non-relativistic case.

In the zero-momentum frame, the magnitudes of momenta of projectiles a and target A in the A(a,b)B reaction, denoted $p_\textrm{a}^{(Z)}$ and $p_\textrm{A}^{(Z)}$, respectively, are equal to their relative momentum $p_\textrm{in}^{(Z)}$, where the subscript ``in'' refers to the incident channel. According to relativistic kinematics, they are given by~\cite{williams1971}
\begin{equation}
p_\mathrm{a}^{(Z)}=p_\mathrm{A}^{(Z)}=p_\mathrm{in}^{(Z)}=
p_{\mathrm{a}}^{\mathrm{(L)}}\sqrt{\frac{m_{\mathrm{A}}^2}{m_{\mathrm{A}}^2+m_{\mathrm{a}}^2+2m_{\mathrm{A}}E_{\mathrm{a}}^{\mathrm{(L)}}}},
\end{equation}
where $p_\mathrm{a}^\mathrm{(L)}$ is the momentum of particle a in the laboratory system, which relates the incident energy $T_\mathrm{in}^\mathrm{(L)}$, the rest mass of particle a ($m_\mathrm{a}$), and the total energy of particle a ($E_\mathrm{a}^\mathrm{(L)}$) by
\begin{equation}
E_\mathrm{a}^\mathrm{(L)}=m_\mathrm{a}+T_\mathrm{in}^\mathrm{(L)}
=\sqrt{m_\mathrm{a}^2+\left(p_\mathrm{a}^\mathrm{(L)}\right)^2}.
\end{equation}
The total energies of particles A and a, $E_\mathrm{A}^\mathrm{(Z)}$ and $E_\mathrm{a}^\mathrm{(Z)}$, and their kinetic energies, $T_\mathrm{A}^\mathrm{(Z)}$ and $T_\mathrm{a}^\mathrm{(Z)}$, respectively, are related to $p_\textrm{in}^{(Z)}$ by
\begin{eqnarray}
E_\mathrm{A}^\mathrm{(Z)}&=&\sqrt{m_\mathrm{A}^2+\left(p_\textrm{in}^{(Z)}\right)^2},\\
E_\mathrm{a}^\mathrm{(Z)}&=&\sqrt{m_\mathrm{a}^2+\left(p_\textrm{in}^{(Z)}\right)^2},\\
T_\mathrm{A}^\mathrm{(Z)}&=& \frac{\left(p_\textrm{in}^{(Z)}\right)^2}{E_\mathrm{A}^\mathrm{(Z)}+m_\mathrm{A}},\\
T_\mathrm{a}^\mathrm{(Z)}&=& \frac{\left(p_\textrm{in}^{(Z)}\right)^2}{E_\mathrm{a}^\mathrm{(Z)}+m_\mathrm{a}},
\end{eqnarray}
and the total kinetic energy in the ZMF is given by
\begin{eqnarray}\label{eq-Tinz}
T_\mathrm{in}^\mathrm{(Z)}
&=&E_{\mathrm{A}}^{\mathrm{(Z)}}+E_{\mathrm{a}}^{\mathrm{(Z)}}-m_{\mathrm{A}}-m_{\mathrm{a}}\\
&=&T_\mathrm{A}^\mathrm{(Z)}+T_\mathrm{a}^\mathrm{(Z)}=
\frac{\left(p_\textrm{in}^{(Z)}\right)^2}{E_\mathrm{A}^\mathrm{(Z)}+m_\mathrm{A}}+
\frac{\left(p_\textrm{in}^{(Z)}\right)^2}{E_\mathrm{a}^\mathrm{(Z)}+m_\mathrm{a}}.\nonumber
\end{eqnarray}

To preserve the same formal relation between the kinetic energy $T_\mathrm{in}^\mathrm{(Z)}$ and the relative momentum $p_\mathrm{in}^\mathrm{(Z)}$ as in the non-relativistic case, namely $2\mu_\mathrm{in}T_\mathrm{in}^\mathrm{(Z)}=\left(p_\mathrm{in}^\mathrm{(Z)}\right)^2$,  the reduced mass in the incident channel must be defined as
\begin{equation}
\mu_\mathrm{in}=\frac{1}{\frac{2}{m_\mathrm{A}+E_\mathrm{A}^\mathrm{(Z)}}+\frac{2}{m_\mathrm{a}+E_\mathrm{a}^\mathrm{(Z)}}},
\end{equation}
which is satisfied by replacing their physical masses $m_\mathrm{A}$ and $m_\mathrm{a}$ with the effective masses
\begin{equation}\label{eq-meffa}
M_\mathrm{eff,A} = \frac{m_\mathrm{A}+E_\mathrm{A}^\mathrm{(Z)}}{2}, \textrm{ and }
M_\mathrm{eff,a} = \frac{m_\mathrm{a}+E_\mathrm{a}^\mathrm{(Z)}}{2}.
\end{equation}
With these replacements, the reduced mass in the zero-momentum frame takes the same functional form as in the non-relativistic case:
\begin{equation}\label{mu_in_redM}
\mu_\mathrm{in}=\frac{M_\mathrm{eff,A}M_\mathrm{eff,a}}{M_\mathrm{eff,A}+M_\mathrm{eff,a}}.
\end{equation}
When these effective masses are used in non-relativistic reaction codes, the center-of-mass energy of the projectile-target system is calculated in the usual manner as
\begin{equation}\label{eq-ecm}
E_\mathrm{c.m.}=\frac{M_{\mathrm{eff},\mathrm{A}}}{M_{\mathrm{eff},\mathrm{A}} + M_{\mathrm{eff},\mathrm{a}}}T_{\mathrm{in}}^{\mathrm{(L)}}.
\end{equation}
This provides an excellent approximation to the total kinetic energy $T_\mathrm{in}^{(Z)}$ in the ZMF (Eq.(\ref{eq-Tinz})). For instance, for proton elastic scattering from \nuc{16}{O} at $T_{\mathrm{in}}^{\mathrm{(L)}}$=1 GeV, the difference between $E_\mathrm{c.m.}$ and $T_\mathrm{in}^{(Z)}$ is less than 0.3\%.
Furthermore, the magnitude of the Lorentz-covariant momentum $p_{\mathrm{in}}^{\mathrm{(Z)}}$ in the ZMF can also be well approximated using the effective masses and the incident energy:
\begin{equation} \label{eq-momentum}
p_{\mathrm{in}}^{\mathrm{(Z)}} = \sqrt{2\frac{M_{\mathrm{eff},\mathrm{A}}M_{\mathrm{eff},\mathrm{a}}}{M_{\mathrm{eff},\mathrm{A}} + M_{\mathrm{eff},\mathrm{a}}} T_{\mathrm{in}}^{\mathrm{(Z)}}}
\approx \frac{M_{\mathrm{eff},\mathrm{A}} \sqrt{2M_{\mathrm{eff},\mathrm{a}}T_{\mathrm{in}}^{\mathrm{(L)}}}}{M_{\mathrm{eff},\mathrm{A}} + M_{\mathrm{eff},\mathrm{a}}}.
\end{equation}

To apply non-relativistic reaction codes within the zero-momentum frame to relativistic cases, the relative momentum $p_\mathrm{out}^{(Z)}$ in the outgoing channel of the A(a,b)B reaction must satisfy the same formal relation as in non-relativistic kinematics: $\left(p_\mathrm{out}^{(Z)}\right)^2=2\mu_\mathrm{out}T_\mathrm{out}^{(Z)}$, where $\mu_\mathrm{out}$ is the reduced mass of B and b, and $T_\mathrm{out}^{(Z)}=T_\mathrm{in}^{(Z)}+Q$ is the total kinetic energy in the outgoing channel, with $Q$ the reaction $Q$-value. To achieve this, the physical masses $m_\mathrm{B}$ and $m_\mathrm{b}$ must be replaced by the effective masses
\begin{equation}\label{eq-meffb}
M_\mathrm{eff,B} = \frac{m_\mathrm{B}+E_\mathrm{B}^\mathrm{(Z)}}{2} \textrm{ and }
M_\mathrm{eff,b} = \frac{m_\mathrm{b}+E_\mathrm{b}^\mathrm{(Z)}}{2},
\end{equation}
where
\begin{equation}
E_\mathrm{B}^\mathrm{(Z)}=\sqrt{m_\mathrm{B}^2+\left(p_\textrm{out}^{(Z)}\right)^2} \textrm{ and }
E_\mathrm{b}^\mathrm{(Z)}=\sqrt{m_\mathrm{b}^2+\left(p_\textrm{out}^{(Z)}\right)^2}, 
\end{equation}
and
\begin{equation}
p_\textrm{out}^{(Z)}=\frac{1}{2E_\mathrm{tot}^{(Z)}}
\sqrt{\left(E_\mathrm{tot}^{(Z)}\right)^4-2\left(m_\mathrm{B}^2+m_\mathrm{b}^2\right)\left(E_\mathrm{tot}^{(Z)}\right)^2+\left(m_\mathrm{B}^2-m_\mathrm{b}^2\right)^2}.
\end{equation}
Here, the total energy in the outgoing channel is given by
\begin{equation}
E_\mathrm{tot}^{(Z)}\equiv E_\mathrm{B}^{(Z)}+E_\mathrm{b}^{(Z)}
=E_\mathrm{A}^{(Z)}+E_\mathrm{a}^{(Z)}
=T_\mathrm{in}^{(Z)}+Q+m_\mathrm{B}+m_\mathrm{b}.
\end{equation}
With the effective masses defined in Eq.(\ref{eq-meffb}), the reduced mass of b and B is given by
\begin{equation}
\mu_\mathrm{out} \equiv \frac{\left(E_{\mathrm{B}}^{\mathrm{(Z)}} + m_{\mathrm{B}}\right)\left(E_{\mathrm{b}}^{\mathrm{(Z)}} + m_{\mathrm{b}}\right)}{ 2\left(E_{\mathrm{B}}^{\mathrm{(Z)}} + m_{\mathrm{B}} + E_{\mathrm{b}}^{\mathrm{(Z)}} + m_{\mathrm{b}}\right)}.
\end{equation}
Since this relativistic kinematic correction scheme requires only the replacement of particle masses according to Eqs.(\ref{eq-meffa}) and (\ref{eq-meffb}), we refer to it as the ``redM'' method in the following discussion.

Another widely used relativistic correction is the one introduced in Ref.~\cite{Satchler-NPA-1992}. Its prescription is to replace the rest masses of the particles with their relativistic ones, i.e., 
\begin{equation}\label{eq-m-redE}
M_\mathrm{eff,i}^\mathrm{redM}=E_\mathrm{i}^{(Z)} \quad \mathrm{(i=A, a, B, b)}
\end{equation}
so that the reduced masses in the incident and outgoing channels are
\begin{equation}\label{mu_in_redE}
\mu_{\mathrm{in,redE}}\equiv\frac{M_\mathrm{eff,A}^\mathrm{redM}M_\mathrm{eff,a}^\mathrm{redM}}{M_\mathrm{eff,A}^\mathrm{redM}+M_\mathrm{eff,a}^\mathrm{redM}}=\frac{E_{\mathrm{A}}^{\mathrm{(Z)}}E_{\mathrm{a}}^{\mathrm{(Z)}}}{E_{\mathrm{A}}^{\mathrm{(Z)}}+E_{\mathrm{a}}^{\mathrm{(Z)}}},
\end{equation}
and
\begin{equation} \mu_{\mathrm{out,redE}}\equiv\frac{M_\mathrm{eff,B}^\mathrm{redM}M_\mathrm{eff,b}^\mathrm{redM}}{M_\mathrm{eff,B}^\mathrm{redM}+M_\mathrm{eff,b}^\mathrm{redM}} =\frac{E_{\mathrm{B}}^{\mathrm{(Z)}}E_{\mathrm{b}}^{\mathrm{(Z)}}}{E_{\mathrm{B}}^{\mathrm{(Z)}}+E_{\mathrm{b}}^{\mathrm{(Z)}}},
\label{redmass2}
\end{equation}
respectively. We refer to it as the ``redE'' method in this work. 
To use this method in usual non-relativistic direct nuclear reaction codes, not only must the nuclear masses be replaced according to Eqs. (\ref{eq-meffa}) and (\ref{eq-meffb}), but the relative energies associated with the distorted waves in the incident and outgoing channels must also be modified accordingly:
\begin{equation}
T_{\mathrm{in,redE}}=\frac{\left(p_{\mathrm{in}}^{\mathrm{(Z)}}\right)^2}{2\mu_{\mathrm{in,redE}}} \textrm{ and }
T_{\mathrm{out,redE}}=\frac{\left(p_{\mathrm{out}}^{\mathrm{(Z)}}\right)^2}{2\mu_{\mathrm{out,redE}}}.
\end{equation}
The effective reaction $Q$-value is then given by
\begin{equation}
Q_\mathrm{eff}=T_{\mathrm{out,redE}}-T_{\mathrm{in,redE}}.
\end{equation}

\begin{figure}[H]
    \centering
    \includegraphics[width=0.8\linewidth]{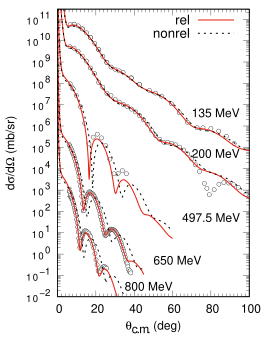}
    \caption{Comparison of theoretical calculations with experimental data for proton elastic scattering from \nuc{16}{O} at various incident proton energies $E_p$ (indicated in the figures). Results calculated with and without relativistic kinematics corrections are shown by the solid and dotted curves, respectively. For clarity, the data sets from top to bottom have been multiplied by factors of 10$^{8}$, 10$^{7}$,10$^{5}$, 10$^{2}$, and 10$^{0}$, respectively. The experimental data are taken from Refs.~\cite{Kelly-PRL-2012-135el,Glover-PRC-1985-200el,Flanders-PRC-1991-497.5el,Bleszynski-PRC-1988-650MeV,Adams-PRL-1979-800el}. See the text for further details.}
    \label{fig-ela-scat}
\end{figure}

Compared with the ``redE'' method, the ``redM'' method requires minimal efforts to incorporate RKCs into existing non-relativistic direct nuclear reaction codes, as it only requires replacing the particle masses according to Eqs. (\ref{eq-meffa}) and (\ref{eq-meffb}). 
It should be noted that both corrections apply only to the calculation of the distorted waves in the incident and outgoing channels; the bound-state wave functions of the $n$+B and $n$+$p$ systems are calculated non-relativistically. In practice, these wave functions are computed separately and read into the code when the mass replacements in Eqs. (\ref{eq-meffa}) and (\ref{eq-meffb}) or in Eq.~(\ref{eq-m-redE}) are employed, so as to avoid unintended modification of the bound-state wave functions due to the altered masses.

Before examining the effects of RKCs on $(p,d)$ reactions, we firstly assess their validity in nucleon-nucleus elastic scattering. 
This also allows us to verify the optical model potentials (OMPs) to be used in the subsequent $(p,d)$ reaction calculations. 
To this end, we study proton elastic scattering from \nuc{16}{O} at incident energies of 135, 200, 497.5, 650, and 800 MeV. 
The results of optical model calculations, performed with and without RKCs, are compared with experimental data in Fig.~\ref{fig-ela-scat}. The results shown are those obtained with the ``redM" method. They are identical to those from the ``redE" method. With these corrections, the elastic scattering angular distributions are described very well. 
The OMPs employed here are the Schr\"odinger-equivalent global Dirac optical model potentials from Ref.~\cite{Cooper-PRC-1993}, which were obtained by fitting proton elastic scattering data on \nuc{12}{C}, \nuc{16}{O}, \nuc{40}{Ca}, \nuc{90}{Zr}, and \nuc{208}{Pb} targets over an energy range of 20-1040 MeV.
The results shown in Fig.~\ref{fig-ela-scat} confirm both the effectiveness of the RKCs and the suitability of these OMPs for the $(p,d)$ reaction calculations presented in the next section.

\begin{figure}[!ht]
    \centering
    \includegraphics[width=\linewidth]{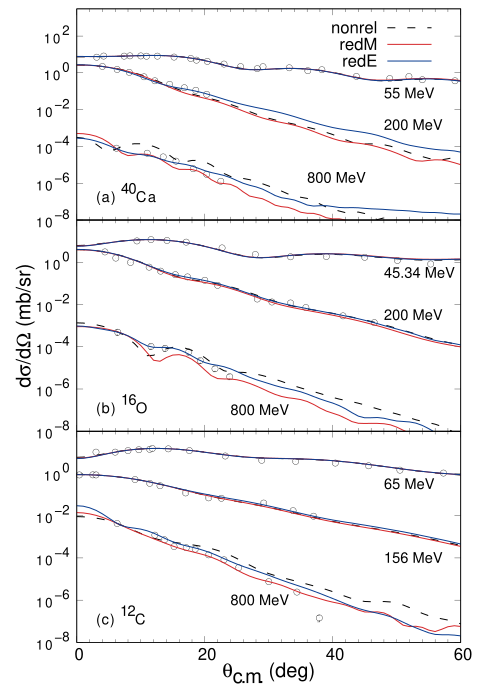}
    \caption{Differential cross sections of the $(p,d)$ reaction on \nuc{40}{Ca}, \nuc{16}{O}, and \nuc{12}{C} at various incident energies, with residual nuclei left in their ground states. Experimental data are shown as hollow circles. Red lines represent ADWA calculations with the ``redM'' relativistic correction, blue lines with the ``redE'' relativistic correction, and dashed lines without any relativistic correction. The incident proton energies are indicated in each subfigure. For clarity, the experimental cross sections and the theoretical calculations for the $^{16}$O($p$,$d$)$^{15}$O  reaction at 200 MeV have been additionally multiplied by a factor of 0.1. 
     The experimental data in subfigure (a) are from \cite{Ejiri-40Capd55MeV-1966,Abegg-40Capd200MeV-PRC-1989,Smith-16Opd800MeV-PRC-1984}; in subfigure (b) from \cite{Snelgrove-16Opd45.34MeV-PR-1969,Abegg-40Capd200MeV-PRC-1989,Smith-16Opd800MeV-PRC-1984}; and in subfigure (c) from \cite{ROOS-12Cpd65MeV-NPA-1975,Bachelier-12Cpd156MeV-NP-1969,Smith-16Opd800MeV-PRC-1984}.}
    \label{fig-cross-sections}
\end{figure}

\begin{figure}
    \centering
    \includegraphics[width=\linewidth]{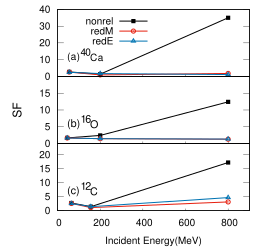}
    \caption{Neutron spectroscopic factors (SFs) of \nuc{40}{Ca}, \nuc{16}{O}, and \nuc{12}{C} extracted from the $(p,d)$ reaction at various incident energies. Black squares represent SFs obtained without relativistic corrections. Red open circles and blue open triangles represent SFs extracted with the ``redM'' and ``redE'' relativistic correction schemes, respectively. The solid lines are drawn to guide the eye.}
    \label{sf}
\end{figure}

To investigate the effects of relativistic corrections on ($p$,$d$) reactions, we study ($p$,$d$) reactions on \nuc{12}{C}, \nuc{16}{O}, and \nuc{40}{Ca} targets, with the residual nuclei left in their ground states, at incident energies from approximately 50 to 800 MeV. Calculations are performed both with and without relativistic corrections.
The proton OMPs used are the same Schr\"odinger-equivalent Dirac optical potentials as those in Fig. \ref{fig-ela-scat}. 
Employing the zero-range adiabatic approximation~\cite{Johnson-PRC-1970}, we construct the deuteron potential by adding the central parts of the proton and neutron potentials, assuming the nuclear parts of the proton-target and neutron-target potentials are identical. The depth of the deuteron spin-orbit potential is taken to be half that of the proton.
There are exceptions for the $(p,d)$ reactions on \nuc{12}{C} at 65 MeV and \nuc{40}{Ca} at 55 MeV, for which the systematic Schr\"odinger-equivalent Dirac potentials fail to describe the incident-channel elastic scattering data sufficiently well. 
In these cases, the phenomenological OMPs of Koning and Delaroche~\cite{KONING-NPA-2003} are used instead. In all calculations, the Reid soft-core nucleon-nucleon interaction~\cite{Reid1968,Stoks1994} is used to generate the deuteron wave functions, which include both $S$- and $D$-wave components. The neutron single-particle wave functions are determined using the usual separation-energy prescription with Woods-Saxon potentials without spin-orbit components. The diffuseness parameter $a_0$ is fixed at 0.65 fm, and the radius parameters $r_0$ for \nuc{40}{Ca}, \nuc{16}{O}, and \nuc{12}{C} are 1.35 fm, 1.43 fm, and 1.33 fm, respectively~\cite{HaiWenlong-PRC-2024}. 
All calculated cross sections are normalized to the experimental data at the center-of-mass angles where the measured cross sections reach their maxima, from which we extract the neutron spectroscopic factors (SFs) for the ground states of the target nuclei.

Results of the ADWA calculations are shown in Fig.~\ref{fig-cross-sections}, and the extracted SFs are plotted as a function of the incident energies in Fig.~\ref{sf}. 
It is evident that relativistic effects gradually manifest as the incident energy increases. At higher energies, the SFs extracted from non-relativistic calculations are appreciably larger than those obtained with relativistic corrections. 
Although the cross sections obtained with the ``redM" and ``redE" methods are not identical, their magnitudes are similar, leading to comparable SFs. Both sets of cross sections for the \nuc{16}{O}($p$,$d$)\nuc{15}{O} at 200 MeV are consistent with the DWBA calculation reported in Ref.~\cite{Shim2026}, with the present results reproducing the experimental data better at larger scattering angles.
These results suggest that the kinematic features of the incident and outgoing channels play a dominant role in determining the cross section magnitude at high energies, while the difference between the two relativistic correction methods arises from the different effective masses employed.
Notably, the relativistic calculations yield SFs that are significantly more consistent across the 50-800 MeV energy range.
In contrast, the substantial variation in SFs extracted from non-relativistic calculations underscores the necessity of including relativistic corrections in high-momentum transfer reaction calculations.

It is important to understand why non-relativistic calculations produce substantially larger SFs. Following the idea of G. F. Chew and M. L. Goldberger~\cite{Chew-PR-1950}, we consider the plane-wave approximation to shed light on this question.
In this approximation, the transition amplitude for the A($p$,$d$)B reaction is given by the product of two Fourier transforms: the first involving the deuteron ground-state wave function $\phi_d(\bm{r})$ and the $p$-$n$ interaction $V_{pn}(\bm{r})$, and the second involving the single-particle wave function $\phi_{n\ell j}(\bm{r}_{n})$ of the transferred neutron in nucleus A~\cite{Glendenning-book}:
\begin{equation}
T_{pd}^{\textrm{PW}} = \int{e^{-i\bm{K}\cdot \bm{r}} \phi_{d}^{\ast}(\bm{r}) V_{pn}(\bm{r}) \, d\bm{r}}\times \int{e^{-i\bm{q}\cdot \bm{r}_n} \phi_{n\ell j} (\bm{r}_n) \, d\bm{r}_n,}\label{eq-plane-wave}
\end{equation}
where $\bm{K}\equiv \bm{k}_p-\bm{k}_d/2$ and $\bm{q}\equiv \bm{k}_d-((m_\mathrm{A}-m_p)/m_\mathrm{A})\bm{k}_p$. 
As discussed in Ref.~\cite{Chew-PR-1950}, the second integral in Eq.~(\ref{eq-plane-wave}) -- that is, the Fourier transform of the single-nucleon wave function -- dominates the calculation. Let us denote this term by $N(\bm{q})$:
\begin{equation}
    N(\bm{q})\equiv \int e^{-i\bm{q}\cdot \bm{r}_n} \phi_{n\ell j}\bigl( \bm{r}_n \bigr) \, d\bm{r}_n.
\end{equation}
Within the plane-wave approximation, the differential cross section for the $(p,d)$ reaction is therefore proportional to $N^2(\bm{q})$.

\begin{figure}
    \centering
    \includegraphics[width=0.8\linewidth]{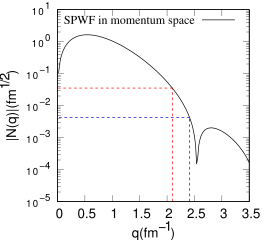}
    \caption{Radial part of the momentum space single-particle wave function of the transferred neutron in the \nuc{12}{C} ($p$,$d$)\nuc{11}{C} reaction, plotted as an absolute value. The red and blue vertical dashed lines indicate the zero-degree momentum transfer for the \nuc{12}{C}($p$,$d$)\nuc{11}{C} reaction at 800 MeV, calculated with and without the RKCs, respectively. The corresponding values of the momentum-space wave function are exactly $N(q)$ at 0$^{\circ}$.}
    \label{fig-spwf}
\end{figure}

As seen in Fig.~\ref{fig-cross-sections}, the differential cross sections of $(p,d)$ reactions peak at zero degrees in the center-of-mass system at high energies. We therefore study the effect of RKCs by examining the differential cross section at zero degrees, where the momentum transfer is
$q_0 \equiv |\bm{k}_d|-((m_\mathrm{A}-m_p)/m_\mathrm{A})|\bm{k}_p|.$
Because the magnitudes ($k_{d}$ and $k_{p}$) differ depending on whether RKC are applied -- as illustrated by Eq.~(\ref{eq-momentum}), where the relativistic momentum is calculated using the same functional form as in the non-relativistic case but with modified masses, the resulting $q_0$ values also differ accordingly. 
This in turn affects $N(q_0)$ and, consequently, the calculated differential cross sections at zero degrees, which are proportional to $N^2(q_0)$. 
As an example, Fig.~\ref{fig-spwf} shows the $N(q)$ values for the single-nucleon wave function in the ground state of \nuc{12}{C}. The corresponding $k_{d}$, $k_{p}$ and $q_0$ values for the \nuc{12}{C}$(p,d)$\nuc{11}{C} reaction, calculated at different incident proton energies with and without RKC, are listed in Table~\ref{tab-K-k}. At 800 MeV, the $q_0$ value obtained with RKC is smaller than that from the non-relativistic calculation. This leads to a larger $N(q_0)$, which increases the zero-degree differential cross section and results in a smaller extracted spectroscopic factor.

\begin{table}[h]
    \centering
    \caption{Values of the incident proton wave number $k_p$, the outgoing deuteron wave number $k_d$, and $q_0$ for the \nuc{12}{C}($p$,$d$)\nuc{11}{C} reaction, calculated without and with relativistic corrections.}
    \begin{tabular}{ccccccc} 
    \toprule
    &\multicolumn{3}{c}{Nonrelativistic}&\multicolumn{3}{c}{Relativistic} \\ 
    \cmidrule{2-4}\cmidrule{5-7}
    \makecell[c]{Energy} &\makecell[c]{$k_p$}&
    \makecell[c]{$k_d$}&\makecell[c]{$q_0$}&
    \makecell[c]{$k_p$}&\makecell[c]{$k_d$}&
    \makecell[c]{$q_0$}\\
    (MeV)&\multicolumn{3}{c}{(fm$^{-1}$)}&\multicolumn{3}{c}{(fm$^{-1}$)} \\ 
    \midrule
 50   &  1.43	 & 1.55 & 0.24    &1.44 &  1.55 & 0.23 \\   
 100  &  2.02	 & 2.48 & 0.63    &2.06 &  2.49 & 0.60 \\
 200  &  2.85	 & 3.69 & 1.07    &2.96 &  3.73 & 1.02 \\
 400  &  4.04	 & 5.34 & 1.64    &4.32 &  5.47 & 1.51 \\
 600  &  4.95	 & 6.60 & 2.06    &5.44 &  6.83 & 1.84 \\
 800  &  5.71	 & 7.65 & 2.41    &6.45 &  8.01 & 2.10 \\
    \bottomrule
    \end{tabular}
    \label{tab-K-k}
\end{table}

In summary, we have presented and examined two relativistic kinematic correction schemes --``redM'' and ``redE''-- for incorporating relativistic effects into the ADWA description of $(p,d)$ reactions within the Schr\"odinger-equation framework. 
We have also provided a practical prescription for implementing these corrections in standard non-relativistic DWBA/ADWA codes. Both methods redefine the reduced mass in the zero-momentum frame to obtain the correct Lorentz-covariant momentum in the incident- and outgoing-channels. 
Validation against proton elastic scattering data on $^{16}$O at incident energies from 65 to 800 MeV confirms that both correction schemes, combined with Schr\"odinger-equivalent Dirac global optical model potentials, accurately reproduce the experimental angular distributions. Applied to $(p,d)$ reactions on $^{12}$C, $^{16}$O and $^{40}$Ca, the relativistic corrections yield neutron spectroscopic factors that are significantly more consistent across the 50--800 MeV energy range than non-relativistic ADWA calculations, which systematically overestimate spectroscopic factors at high incident energies.
A plane-wave analysis suggests that this overestimation arises because the non-relativistic zero-degree momentum transfer $q_0$ is larger than its relativistic counterpart. This leads to the sampling of an incorrect Fourier component of the bound-state wave function.

The present approach is limited to kinematic relativistic corrections; dynamic relativistic effects are not addressed. Nevertheless, the consistency of the extracted spectroscopic factors across a wide energy range demonstrates that kinematic relativistic corrections to ADWA are essential for high-energy ($p$,$d$) reactions, and provide a practical route for studying high-momentum components of nuclear wave functions with hadronic probes. 
Residual inconsistencies between low- and high-energy data remain, potentially arising from optical model uncertainties, neglected dynamical corrections, and -- most critically -- the bound-state wave functions used in the reaction calculation whose high-momentum tails may not accurately represent the true neutron momentum distribution. Achieving fully energy-independent spectroscopic factors will therefore require both refined reaction models and more realistic bound-state wave functions.

\section*{Acknowledgements}
This work was supported by the National Key R\&D Program of China (Grant No. 2023YFA1606702), the National Natural Science Foundation of China under Contracts Nos. 12175009, 12175280, 12250610193, 11605253 and 12275007), the International Partnership Program of the Chinese Academy of Sciences (016GJHZ2023063GC), and Major Science and Technology Projects in Gansu Province (24ZD13GA005).
H. J. O. thanks K. Ogata and T. Uesaka for fruitful discussion, and acknowledges the support of CAS President's International Fellowship Initiative.

\bibliographystyle{elsarticle-num} 
\bibliography{RCADWA}

@ARTICLE{Tanihata-MPLA-2010,
	author = {I. Tanihata},
	title = {Searching for effects of tensor forces in nuclei},
	year = {2010},
	journal = {Mod. Phys. Lett. A},
	volume = {25},
	number = {21-23},
	pages = {1886 – 1890},
	doi = {10.1142/S0217732310000563},
	issn = {02177323},
	coden = {MPLAE}
}

@article{Johnson-aip-2005,
    author = {Johnson, R. C.},
    title = {Adiabatic approximation for nucleus‐nucleus scattering},
    journal = {AIP Conference Proceedings},
    volume = {791},
    number = {1},
    pages = {128-139},
    year = {2005},
    month = {10},
    issn = {0094-243X},
    doi = {10.1063/1.2114701},
    url = {https://doi.org/10.1063/1.2114701}
}

@article{Timofeyuk-ppnp-2020,
author = {N. K. Timofeyuk and R. C. Johnson},
title = {Theory of deuteron stripping and pick-up reactions for nuclear structure studies},
journal = {Progress in Particle and Nuclear Physics},
volume = {111},
pages = {103738},
year = {2020},
issn = {0146-6410},
doi = {https://doi.org/10.1016/j.ppnp.2019.103738}
}

@article{Ong2013,
  author = {H. J. Ong and I. Tanihata and A. Tamii and others},
  title = {Probing effect of tensor interactions in {$^{16}$O} via ($p$,$d$) reaction},
  journal = {Phys. Lett. B},
  volume = {725},
  number = {4--5},
  pages = {277--281},
  year = {2013},
  doi = {10.1016/j.physletb.2013.07.038},
}

@article{Terashima2018,
  author = {S. Terashima and L. Yu and H. J. Ong and others},
  title = {Dominance of Tensor Correlations in High-Momentum Nucleon Pairs Studied by ($p$,$pd$) Reaction},
  journal = {Phys. Rev. Lett.},
  volume = {121},
  number = {24},
  pages = {242501},
  year = {2018},
  doi = {10.1103/PhysRevLett.121.242501}
}

@article{Wang2026,
  author = {X. Wang and H. J. Ong and S. Terashima and others},
  title = {Observation of Tensor-Driven High-Momentum Neutrons in {$^{16}$O} via ($p$,$d$) Reactions and Zero-Degree Deuteron Momentum Spectroscopy},
  journal = {Phys. Lett. B},
  volume = {879},
  number = {},
  pages = {140563},
  year = {2026},
  doi = {10.1016/j.physletb.2026.140563},
}

@article{Haiwenlong-PRC-2024,
  title = {Determining the radii of single-particle potentials with {Skyrme Hartree-Fock} calculations},
  author = {W. L. Hai and D. Y. Pang and X. B. Wang and others},
  journal = {Phys. Rev. C},
  volume = {110},
  issue = {4},
  pages = {044613},
  numpages = {12},
  year = {2024},
  month = {Oct},
  publisher = {American Physical Society},
  doi = {10.1103/PhysRevC.110.044613}
 }

@article{Suda2017,
  author = {T. Suda and H. Simon},
  title = {Prospects for electron scattering on unstable, exotic nuclei},
  journal = {Prog. Part. Nucl. Phys.},
  volume = {96},
  number = {},
  pages = {1},
  year = {2017},
  doi = {10.1016/j.ppnp.2017.04.002}
}

@book{Glendenning-book,
author = {Norman K Glendenning},
title = {Direct Nuclear Reactions},
publisher = {World Scientific},
year = {2004},
doi = {10.1142/5612},
address = {},
edition   = {}
}

@article{Subedi2008,
author = {R. Subedi  and R. Shneor  and P. Monaghan and others},
title = {Probing Cold Dense Nuclear Matter},
journal = {Science},
volume = {320},
number = {5882},
pages = {1476-1478},
year = {2008},
doi = {10.1126/science.1156675},
}

@article{Hen-Science-2014,
author = {O. Hen  and M. Sargsian  and L. B. Weinstein  and others},
title = {Momentum sharing in imbalanced Fermi systems},
journal = {Science},
volume = {346},
number = {6209},
pages = {614-617},
year = {2014},
doi = {10.1126/science.1256785}
}

@ARTICLE{CLAS-nature-2018,
	author = {M. Duer and O. Hen and E. Piasetzky and others},
	title = {Probing high-momentum protons and neutrons in neutron-rich nuclei},
	year = {2018},
	journal = {Nature},
	volume = {560},
	pages = {617--621},
	doi = {10.1038/s41586-018-0400-z}
}

@article{Mardor1998,
  author = {Y. Mardor and J. Aclander and J. Alster and others},
  title = {Measurement of quasi-elastic {$^{12}$C($p$,$2p$)} scattering at high momentum transfer},
  journal = {Phys. Lett. B},
  volume = {437},
  number = {3--4},
  pages = {257--263},
  year = {1998},
  doi = {10.1016/S0370-2693(98)01000-4}
}

@article{Tang2003,
  author = {A. Tang and J. W. Watson and J. Aclander and others},
  title = {$n$-$p$ Short-Range Correlations from ($p$,$2p$+$n$) Measurements},
  journal = {Phys. Rev. Lett.},
  volume = {90},
  number = {4},
  pages = {042301},
  year = {2003},
  doi = {10.1103/PhysRevLett.90.042301}
}

@article{Patsyuk2021,
  author = {M. Patsyuk and J. Kahlbow and G. Laskaris and others},
  title = {Unperturbed inverse kinematics nucleon knockout measurements with a carbon beam},
  journal = {Nat. Phys.},
  volume = {17},
  pages = {693},
  year = {2021},
  doi = {10.1038/s41567-021-01193-4}
}

@article{Reid1968,
  author = {R. V. {Reid Jr}},
  title = {Local phenomenological nucleon-nucleon potentials},
  journal = {Ann. Phys.},
  volume = {50},
  pages = {411},
  year = {1968},
  doi = {10.1016/0003-4916(68)90126-7}
}

@article{Stoks1994,
  author = {V. G. J. Stoks and R. A. M. Klomp and C. P. F. Terheggen and J. J. de Swart},
  title = {Construction of high-quality {$NN$} potential models},
  journal = {Phys. Rev. C},
  volume = {49},
  pages = {2950},
  year = {1994},
  doi = {10.1103/PhysRevC.49.2950}
}

@article{Arrington-PPNP-2012,
author = {J. Arrington and D. W. Higinbotham and G. Rosner and M. Sargsian},
title = {Hard probes of short-range nucleon–nucleon correlations},
journal = {Progress in Particle and Nuclear Physics},
volume = {67},
number = {4},
pages = {898-938},
year = {2012},
issn = {0146-6410},
doi = {10.1016/j.ppnp.2012.04.002},
}

@article{Schiavilla-PRL-2007,
  title = {Tensor Forces and the Ground-State Structure of Nuclei},
  author = {R. Schiavilla and R. B. Wiringa and Steven C. Pieper and J. Carlson},
  journal = {Phys. Rev. Lett.},
  volume = {98},
  issue = {13},
  pages = {132501},
  numpages = {4},
  year = {2007},
  month = {Mar},
  publisher = {American Physical Society},
  doi = {10.1103/PhysRevLett.98.132501}
}

@article{Roth-PPNP-2010,
author = {R. Roth and T. Neff and H. Feldmeier},
title = {Nuclear structure in the framework of the Unitary Correlation Operator Method},
journal = {Progress in Particle and Nuclear Physics},
volume = {65},
number = {1},
pages = {50-93},
year = {2010},
issn = {0146-6410},
doi = {10.1016/j.ppnp.2010.02.003},
}

@article{Satchler-NPA-1992,
author = {G. R. Satchler},
title = {Local potential model for pion-nucleus scattering and $\pi^+\pi^-$ excitation ratios},
journal = {Nuclear Physics A},
volume = {540},
number = {3},
pages = {533-576},
year = {1992},
issn = {0375-9474},
doi = {https://doi.org/10.1016/0375-9474(92)90173-H},
}

@article{Pang-PRC-2009,
  author = {D. Y. Pang and P. Roussel-Chomaz and H. Savajols and others},
  title = {Global optical model potential for $A=3$ projectiles},
  journal = {Phys. Rev. C},
  volume = {79},
  issue = {2},
  pages = {024615},
  numpages = {21},
  year = {2009},
  month = {Feb},
  publisher = {American Physical Society},
  doi = {10.1103/PhysRevC.79.024615}
}

@article{Pang-CPC-2014,
author = {D. Y. Pang},
title = {Effects of relativistic kinematics in heavy ion elastic scattering},
journal = {Chinese Physics C},
doi = {10.1088/1674-1137/38/2/024104},
year = {2014},
month = {feb},
volume = {38},
number = {2},
pages = {024104},
}

@Article{KongWJ-NST-2023,
author={Wei-Jia Kong and Dan-Yang Pang},
title={Theoretical uncertainties of {($d$,{$^3$He}) and ({$^3$He},$d$)} reactions owing to the uncertainties of optical model potentials},
journal={Nuclear Science and Techniques},
year={2023},
month={Jun},
day={26},
volume={34},
number={6},
pages={95},
doi={10.1007/s41365-023-01242-y}
}

@Article{Schmidt-nature-2020,
author={A. Schmidt and J. R. Pybus and R. Weiss and others},
title={Probing the core of the strong nuclear interaction},
journal={Nature},
year={2020},
month={Feb},
day={01},
volume={578},
number={7796},
pages={540-544},
doi={10.1038/s41586-020-2021-6}
}

@article{Kyushu,
  title = {},
  author = {M. Yahiro and Y. Iseri and K. Ogata},
  journal = {private communication},
  volume = {},
}

@article{Ciofi-PRC-1996,
  title = {Realistic model of the nucleon spectral function in few- and many-nucleon systems},
  author = {Ciofi degli Atti, C. and Simula, S.},
  journal = {Phys. Rev. C},
  volume = {53},
  issue = {4},
  pages = {1689--1710},
  numpages = {0},
  year = {1996},
  month = {Apr},
  publisher = {American Physical Society},
  doi = {10.1103/PhysRevC.53.1689}
}

@ARTICLE{Ingemarsson-PC-1974,
	author = {Ingemarsson, A.},
	title = {Some notes on optical model calculations at medium energies},
	year = {1974},
	journal = {Physica Scripta},
	volume = {9},
	number = {3},
	pages = {156 – 160},
	doi = {10.1088/0031-8949/9/3/004},
	issn = {00318949}
}

@ARTICLE{Smith-16Opd800MeV-PRC-1984,
	author = {G. R. Smith and J. R. Shepard and R. L. Boudrie and others},
	title = {($p$,$d$) reaction at 800 {MeV}},
	year = {1984},
	journal = {Physical Review C},
	volume = {30},
	number = {2},
	pages = {593 – 615},
	doi = {10.1103/PhysRevC.30.593},
	issn = {05562813}
}

@ARTICLE{Cooper-PRC-1993,
	author = {E. D. Cooper and S. Hama and B. C. Clark and R. L. Mercer},
	title = {Global Dirac phenomenology for proton-nucleus elastic scattering},
	year = {1993},
	journal = {Physical Review C},
	volume = {47},
	number = {1},
	pages = {297 – 311},
	doi = {10.1103/PhysRevC.47.297},
	issn = {05562813}
}

@book{Austern-book,
	title={Direct Nuclear Reaction Theories},
	author={Norman Austern},
	isbn={0-471-03770-2},
	year={1970},
	publisher={World Scientific Publishing Co. Pte. Ltd.}
}

@ARTICLE{Chew-PR-1950,
	author = {G. F. Chew and M. L. Goldberger},
	title = {The production of fast deuterons in high energy nuclear reactions},
	year = {1950},
	journal = {Physical Review},
	volume = {77},
	number = {4},
	pages = {470 – 475},
	doi = {10.1103/PhysRev.77.470},
	issn = {0031899X}
}

@article{Bachelier-12Cpd156MeV-NP-1969,
  author={D. Bachelier and M. Bernas and I. Brissaud and others},
  title={R{\'e}action ($p$,$d$) {\`a} 156 {MeV} et structure des noyaux l{\'e}gers},
  journal={Nuclear Physics},
  year={1969},
  volume={126},
  pages={60-96},
  url={https://api.semanticscholar.org/CorpusID:121592229}
}

@article{ROOS-12Cpd65MeV-NPA-1975,
author = {P. G. Roos and S. M. Smith and V. K. C. Cheng and others},
title = {The ($p$,$d$) reaction at 65 {MeV}},
journal = {Nuclear Physics A},
volume = {255},
number = {1},
pages = {187-203},
year = {1975},
issn = {0375-9474},
doi = {10.1016/0375-9474(75)90157-8},
}

@ARTICLE{Snelgrove-16Opd45.34MeV-PR-1969,
	author = {J. L. Snelgrove and E. Kashy},
	title = {Energy dependence and spectroscopy of the {O16($p$,$d$)O15} reaction},
	year = {1969},
	journal = {Physical Review},
	volume = {187},
	number = {4},
	pages = {1246 – 1258},
	doi = {10.1103/PhysRev.187.1246},
	issn = {0031899X}
}

@article{Ejiri-40Capd55MeV-1966,
  title={The Hole State by the Reactions {$^{40}$Ca($p$,$d$)$^{39}$Ca, $^{52}$Cr($p$,$d$)$^{51}$Cr and $^{60}$Ni($p$, $d$)$^{59}$Ni}},
  author={H. Ejiri and Y. Saji and Y. Ishizaki and others},
  journal={Journal of the Physical Society of Japan},
  year={1966},
  volume={21},
  pages={14-24},
  url={https://api.semanticscholar.org/CorpusID:119031215}
}

@ARTICLE{Abegg-40Capd200MeV-PRC-1989,
	author = {R. Abegg and D. A. Hutcheon and C. A. Miller and others},
	title = {Cross section and analyzing power measurements for the ($p$,$d$) reaction on {O16} and {Ca40} at 200 {MeV}},
	year = {1989},
	journal = {Physical Review C},
	volume = {39},
	number = {1},
	pages = {65 – 69},
	doi = {10.1103/PhysRevC.39.65},
	issn = {05562813}
}

@article{KONING-NPA-2003,
author = {A. J. Koning and J.P. Delaroche},
title = {Local and global nucleon optical models from 1 {keV} to 200 {MeV}},
journal = {Nuclear Physics A},
volume = {713},
number = {3},
pages = {231-310},
year = {2003},
issn = {0375-9474},
doi = {10.1016/S0375-9474(02)01321-0},
}

@ARTICLE{Johnson-PRC-1970,
	author = {R. C. Johnson and P. J. R. Soper},
	title = {Contribution of deuteron breakup channels to deuteron stripping and elastic scattering},
	year = {1970},
	journal = {Physical Review C},
	volume = {1},
	number = {3},
	pages = {976 – 990},
	doi = {10.1103/PhysRevC.1.976},
	issn = {05562813}
}

@article{Schiffer-PRL-2012,
  title = {Test of Sum Rules in Nucleon Transfer Reactions},
  author = {J. P. Schiffer and C. R. Hoffman and B. P. Kay and others},
  journal = {Phys. Rev. Lett.},
  volume = {108},
  issue = {2},
  pages = {022501},
  numpages = {5},
  year = {2012},
  month = {Jan},
  publisher = {American Physical Society},
  doi = {10.1103/PhysRevLett.108.022501}
}

@ARTICLE{Liu-NST-2020,
	author = {Wei Liu and Jian-Ling Lou and Yan-Lin Ye and Dan-Yang Pang},
	title = {Experimental study of intruder components in light neutron-rich nuclei via single-nucleon transfer reaction},
	year = {2020},
	journal = {Nuclear Science and Techniques},
	volume = {31},
	number = {2},
	doi = {10.1007/s41365-020-0731-y},
	issn = {10018042},
	coden = {NSETE}
}

@ARTICLE{Sick-PPNP-2007,
	author = {Sick, Ingo},
	title = {Correlations in nuclei},
	year = {2007},
	journal = {Progress in Particle and Nuclear Physics},
	volume = {59},
	number = {1},
	pages = {447 – 454},
	doi = {10.1016/j.ppnp.2007.01.014},
	issn = {01466410},
	coden = {PPNPD}
}

@ARTICLE{Sick-PRL-1980,
	author = {I. Sick and D. Day and J. S. McCarthy},
	title = {Nuclear high-momentum components and y scaling in electron scattering},
	year = {1980},
	journal = {Physical Review Letters},
	volume = {45},
	number = {11},
	pages = {871 – 874},
	doi = {10.1103/PhysRevLett.45.871},
	issn = {00319007}
}

@inproceedings{Yabana2003StructureAR,
  title={Structure and Reactions of Light Exotic Nuclei},
  author={Kazuhiro Yabana and K{\'a}lm{\'a}n Varga and Rezső G. Lovas and Yasuyuki Suzuki},
  publisher={Taylor $\&$ Francis Inc.},
  year={2003},
  doi={10.1201/9780203168271}
}

@article{Kelly-PRL-2012-135el,
  title = {Signatures of Density Dependence in the Two-Nucleon Effective Interaction near 150 {MeV}},
  author = {J. Kelly and W. Bertozzi and T. N. Buti and others},
  journal = {Phys. Rev. Lett.},
  volume = {45},
  issue = {25},
  pages = {2012--2015},
  numpages = {0},
  year = {1980},
  month = {Dec},
  publisher = {American Physical Society},
  doi = {10.1103/PhysRevLett.45.2012}
}

@article{Glover-PRC-1985-200el,
  title = {Optical model analysis of 200 {MeV} {$\vec{p}+^{16}$O} elastic scattering data measured to large momentum transfers},
  author = {C. W. Glover and P. Schwandt and H. O. Meyer and others},
  journal = {Phys. Rev. C},
  volume = {31},
  issue = {1},
  pages = {1--11},
  numpages = {0},
  year = {1985},
  month = {Jan},
  publisher = {American Physical Society},
  doi = {10.1103/PhysRevC.31.1}
}

@article{Flanders-PRC-1991-497.5el,
  title = {Empirical density-dependent effective interaction for nucleon-nucleus scattering at 500 {MeV}},
  author = {B. S. Flanders and J. J. Kelly and H. Seifert and others},
  journal = {Phys. Rev. C},
  volume = {43},
  issue = {5},
  pages = {2103--2126},
  numpages = {0},
  year = {1991},
  month = {May},
  publisher = {American Physical Society},
  doi = {10.1103/PhysRevC.43.2103}
}

@article{Bleszynski-PRC-1988-650MeV,
  title = {Energy dependence of relativistic effects in the elastic scattering of polarized protons from $^{16}\mathrm{O}$ and $^{40}\mathrm{Ca}$},
  author = {E. Bleszynski and B. Aas and D. Adams and others},
  journal = {Phys. Rev. C},
  volume = {37},
  issue = {4},
  pages = {1527--1536},
  numpages = {0},
  year = {1988},
  month = {Apr},
  publisher = {American Physical Society},
  doi = {10.1103/PhysRevC.37.1527}
}

@article{Adams-PRL-1979-800el,
  title = {Microscopic Description of {800-MeV} Polarized-Proton Scattering from $^{16}\mathrm{O}$},
  author = {G. S. Adams and Th. S. Bauer and G. Igo and others},
  journal = {Phys. Rev. Lett.},
  volume = {43},
  issue = {6},
  pages = {421--424},
  numpages = {0},
  year = {1979},
  month = {Aug},
  publisher = {American Physical Society},
  doi = {10.1103/PhysRevLett.43.421}
}

@article{Shim2026,
  title = {Description of nucleon transfer reactions at intermediate energies within the impulse approach},
  author = {S. I. Shim and Y. Chazono and K. Yoshida and others},
  journal = {Phys. Rev. C},
  volume = {113},
  issue = {},
  pages = {014613},
  numpages = {0},
  year = {2026},
  month = {},
  publisher = {American Physical Society},
  doi = {10.1103/v75y-1t3x}
}

@book{williams1971,
    title     = {An Introduction to Elementary Particles},
    editor    = {Williams, W. S. C.},
    series    = {Pure and Applied Physics},
    volume    = {12},
    publisher = {Academic Press},
    address   = {New York},
    year      = {1971},
    edition   = {1}
  }






\end{document}